\documentclass[prl,aps,onecolumn,psfig,preprintnumbers,amsmath,amssymb,showpacs]{revtex4}
\usepackage{graphicx}\usepackage{color}

\begin{document}

\title{Finite size effects on the phase diagram of the thermodynamical cluster model}
\author{ S.Mallik$^1$,  F. Gulminelli$^2$,G. Chaudhuri$^1$}

\address{$^1$Theoretical Physics Division, Variable Energy Cyclotron Centre, 1/AF Bidhan Nagar, Kolkata700064,India}

\address{$^2$LPC Caen IN2P3-CNRS/EnsiCaen et Universite, Caen, France}
\begin{abstract}
The thermodynamical cluster model is known to present a first-order liquid-gas phase transition in the idealized case of an uncharged, infinitely
extended medium. However, in most practical applications of this model, the system is finite and charged.
In this paper we study how the phase diagram is modified by finite size and Coulomb effects.
 We show that the thermodynamic anomalies which are associated to the finite system counterpart of first order phase transitions,  are correctly reproduced by this effective model. However, approximations in the calculation of the grandcanonical partition sum prevent obtaining the exact mapping between statistical ensembles which should be associated to finite systems.
The ensemble inequivalence associated to the transition persists in the presence of Coulomb, but the phase diagram is deeply modified with respect to the simple liquid-gas phase transition characteristic of the neutral system.
\end{abstract}

\maketitle

\section{introduction}

In dilute media governed by a hard core repulsion and a short range attractive interaction, it is reasonable to assume that correlations are largely exhausted by clusterization\cite{frenkel,band,fisher}. As a consequence, cluster models appear extremely successful in quantitatively describing systems as different as simple fluids, atomic and molecular clusters, metallic alloys,  nuclear multifragmentation, and neutron star crusts\cite{clusters,fabienne,subal,bondorf,hempel,buyu,furu,noi}. The thermodynamics of such models is very well known at the thermodynamic limit\cite{gargi07,fisher_noi}, but  very few formal studies exist for finite systems, even if these latter are a primary field of application of such models.

 In particular,  exactly solvable microscopic models have been shown to exhibit thermodynamic anomalies\cite{cneg_exact}. When a first order phase transition exists at the thermodynamical limit, the microcanonical entropy of the corresponding finite system presents a positive convexity, which is forbidden at the thermodynamic limit and leads to a negative heat capacity in the microcanonical ensemble\cite{gross}. This concept was extended to other statistical ensembles\cite{lattice_prl}.
A negative susceptibility should be observed if  the order parameter is a conserved quantity, while in the statistical ensemble where the order parameter is fixed by an external field, the phase transition is signalled by bimodal distributions of observables correlated to the order parameter\cite{lattice_prl}.

A case of particular interest concerns frustrated systems, because the long range Coulombic interaction is known  to quench the phase transition of simple fluids\cite{grousson}, a phenomenology which is notably at play in the astrophysical context of neutron star crusts\cite{pasta}.

Concerning cluster models, negative susceptibilities have been reported both in the canonical\cite{gargi_fra} and in the Monte-Carlo version of the microcanonical ensemble\cite{raduta} if clusters are electrically neutral.
In both cases it was shown that phase transition signals are affected
by the presence of the Coulomb interaction, and the liquid-gas phase transition is quenched for heavily charged systems.
However, in these works the phase diagram of the model was not computed and it is not clear if the fragmentation phenomenon under strong Coulomb fields can still be viewed as a manifestation of the liquid-gas phase transition\cite{gargi_fra,our_plb}.

In ref.\cite{lehaut}, the phase diagram of the finite charged three-dimensional Ising model was computed. It was reported that the introduction of the Coulomb interaction leads to a deep modification of the phase diagram, where the liquid-gas phase transition is replaced by a different, Coulomb induced, first-order fragmentation transition at lower temperature, similar to the phenomenon of nuclear fission.
Based on these findings, it was suggested that the phenomenon of nuclear multifragmentation could be associated to this Coulomb-driven transition.  However this connection could only be qualitative, because the classical Lattice gas model cannot quantitatively describe nuclear data.
It is therefore interesting to see if a similar phenomenology is present in the cluster model, whose predictive power in the description of fragmentation data was clearly proved over the past decades\cite{bondorf,subal}.

Based on these motivations, in this paper we study the phase diagram of the finite cluster model, and its modifications  in the presence of the Coulomb interaction.

We show that approximations in the calculation of the grandcanonical partition sum prevent to recover the analytical relationships among statistical ensembles which should hold in finite systems.
In spite of that, the thermodynamic anomalies which have been reported in the finite system counterpart of first order phase transitions are correctly reproduced by such effective model when neutral clusters are considered.
In a finite system density can be varied both by varying the number of particles inside a fixed volume, or varying the volume of the box in which a fixed number of particles is contained. Only this latter situation is relevant for the description of the fragmentation phase transition observed in nuclear collisions. We show in detail that these two physical situations lead to deeply different phase diagrams, though in both cases a liquid-gas phase transition is observed.
In the presence of a sufficiently strong Coulomb interaction, the liquid-gas phase transition is quenched, in agreement with previous studies. This suggests that the transition might be better observable in nuclear fragmentation of light systems. However a marginal region of ensemble inequivalence  persists at low temperature and high density, related to  a transition between a dilute gas phase and a phase of finite size droplets. Interesting enough, a backbending in the $P(\rho)$ equation of state, which is maybe the most intuitive signal proposed for  the nuclear liquid-gas phase transition, systematically observed in all mean-field models, is only observed by varying the number of particles in a fixed volume, while in the physical situation relevant for nuclear fragmentation there is no upraising branch at high density.

\section{The (Grand Canonical) Canonical Thermodynamical Model}
As we have stated in the introduction, the basic principle of the thermodynamic cluster model is the hypothesis\cite{fisher} that in dilute media governed by short range attractive forces, inter-particle interactions essentially lead to the formation of clusters. The system of interacting particles is thus modelized as an ideal gas of non-interacting clusters.

For simplicity we assume a simple fluid of stuctureless particles of mass $m$, though cluster models of binary fluids
and binary alloys have been developed for different applications\cite{clusters,fabienne,subal}.

Assuming that the ground state energy $E_0(s)$  of a cluster of size $s$ and mass $M(s)=sm+E_0(s)$ is known as well as its full spectrum of excited states $E_i(s)$,  the partition sum of an isolated cluster at inverse temperature $\beta=T^{-1}$
in a volume $V$  reads

\begin{equation}
\omega_s=\frac{V}{h^3} \left ( \frac{2\pi m s}{\beta} \right )^{3/2}
\cdot \sum_i g_i(s) \exp\left [ -\beta  E_i(s)\right ]= \frac{V}{h^3} \left ( \frac{2\pi m s}{\beta} \right )^{3/2}
\cdot  \exp\left [ -\beta F_\beta(s)) \right] , \label{omegas}
\end{equation}

where $g_i$ is the degeneracy of state $i$,  the cluster entropy is given by
$\exp S(s,e)=\sum_i g_i(s) \delta(e-E_i(s))$, and we have made a saddle point approximation,
\begin{equation}
\int_{-\infty}^{\infty} de \exp \left[ S(s,e) - \beta e \right] \approx exp[-\beta F_\beta(s)] ,
\label{saddle}
\end{equation}
where the cluster free energy   is given  by
$F_{\beta}=\langle E(s) \rangle_{\beta} -T \langle S(s) \rangle_{\beta}$.

An exact calculation of the cluster energy states from the microscopic hamiltonian is out of scope, especially
if the constituent particles have to be treated within quantum mechanics, as it is the case for instance for nuclear clusters. However, mean-field or density functional methods can be employed.
In this paper we will use a standard phenomenological prescription\cite{frenkel,fisher} consisting of a volume and a surface term
\begin{equation}
\langle F(s) \rangle_{\beta}=F_b(\beta,s)s+F_s(\beta,s)s^{2/3}
\end{equation}
appropriate for compact clusters in three dimensions.
To optimize the predictive power for nuclear clusters, we choose the temperature dependence appropriate for
fermion systems:
\begin{equation}
F_b = W_0 +\epsilon_0^{-1} \beta^{-2} \; ; \; F_s= -\sigma(\beta)
\label{fneutral}
\end{equation}
and fix the numerical value of the parameters such as to reproduce nuclear phenomenology:  $W_0=15.8 MeV$, $\epsilon_0=16.0$ MeV and $\sigma(\beta)=\sigma_{0}\{(\beta_{c}^{-2}-\beta^{-2})/(\beta_{c}^{-2}+\beta^{-2}\}^{5/4}$ where $\sigma_{0}=18.0$ MeV and $\beta_{c}=1/18.0 MeV^{-1}$.

An important point in the definition of a cluster model concerns the definition of the cluster self-volume. Early applications considered point-like clusters\cite{frenkel,band}.
The simplest implementation of a finite self-volume consists in assuming completely incompressible clusters with a radius  $R_s\propto s^{1/3}$ in three dimensions. Within this description, the volume appearing in eq.(\ref{omegas}) has to be reduced from total volume ($V_{tot}$) by the volume occupied by the clusters. This is readily calculated if the density is homogeneous (compact clusters): $V=V_{tot}-V_0(N)$, where  $V_0$ represents the volume which would be occupied if all of the $N$ particles would make a single cluster. In this paper we will adopt this description and discuss in detail the important implications that this excluded volume mechanism induces in the grandcanonical thermodynamics. Specifically, we will show that the grandcanonical partition sum cannot be analytically calculated for finite systems in the presence of excluded volume.
An interesting extension of the cluster model to compressible clusters can be found in refs.
\cite{bugaev,esym}.


Eq.(\ref{fneutral}) describes charge neutral clusters. The introduction of electric charge leads to an inter-fragment interaction  which in principle breaks the ideal gas free energy additivity. It is however very well known that a good approximation of the Coulomb energy of a system of uniformly charged spherical clusters   is given by the so-called Wigner-Seitz approximation\cite{bondorf}:
\begin{equation}
E_C=\frac{3}{5} \frac{N^2 e^2}{4R} + \sum_{s=2}^{s_{max}(V)} n_s E_s \label{ecoul}
\end{equation}
where $n_s$ is the number of clusters of size $s$,  $R=(3V/4\pi)^{1/3}$, $s_{max}$ is the maximum cluster size allowed in the total volume  $V$, and we have assumed for simplicity that all clusters are charge symmetric, with an effective charge $q=es/2$.
The energies appearing in the second term read:
\begin{equation}
E_s=\frac{3}{5} \frac{s^2 e^2}{4R_s} \left ( 1-\left (\frac{V_0}{V_{tot}} \right )^{1/3} \right )
\end{equation}
We can see that the exact result for the Coulomb energy is recovered in the limit $V_{tot}\to\infty$, where it converges to the sum of the clusters Coulomb energies, and in the limit $V_{tot}\to V_0$, where it gives the energy of a single cluster occupying the whole volume.

The free energy expression (\ref{fneutral}) with the inclusion of a Coulomb energy term in the form (\ref{ecoul}) was successfully employed in the past in the description of nuclear multifragmentation\cite{subal}.  For an application to the inner crust of neutron star and core-collapse supernova matter\cite{hempel,noi} the charge screening from the presence of a uniform electron background has to be accounted for. This can also be readily implemented in the Wigner-Seitz approximation\cite{buyu}, and gives a free energy functional intermediate between the physical case of neutral and charged clusters. In this paper, we will limit ourselves to the two extreme cases of neutral and fully charged systems, and will not consider applications requiring electron screening.

\section{Mapping the different ensembles}

The canonical partition function for a system of $N$ particles at a given inverse temperature $\beta=T^{-1}$ is given
by
\begin{eqnarray}
Q_{N}=\sum\prod \frac{\omega_{s}^{n_{s}}}{n_{s}!}
\end{eqnarray}
Here the sum is over all possible cluster partitions
which satisfy the conservation laws, and $n_{s}$ is
the number of the composites of size $s$ in the given partition.

The average number of clusters with $s$ particles is
seen easily from the above equation to be
\begin{eqnarray}
\langle n_{s}\rangle=\omega_{s}\frac{Q_{N-s}}{Q_{N}}
\end{eqnarray}
The canonical particle number conservation constraint $N=\sum s\times n_{s}$
can be used to obtain a recursion relation
for the partition function
\begin{eqnarray}
Q_{N}=\frac{1}{N}\sum_{s}s\omega_{s}Q_{N-s}
\end{eqnarray}

This recursion relation allows computing all different observables by successive partial derivatives, using standard statistical mechanics expressions.
Specifically, the pressure, chemical potential and mean energy are given by:
\begin{eqnarray}
 \beta p_{\beta}(N,V) &=& \frac{\partial \ln Q_{N}}{ \partial V} |_{N,\beta}  \label{pc1} \\
 \beta \mu_{\beta}(N,V) &=&- \frac{\partial \ln Q_{N}}{ \partial N} |_{V,\beta}  \label{mucan}\\
E_{\beta}(N,V)&=&-\frac{\partial \ln Q_{N}}{\partial \beta } |_{NV}  ,
\end{eqnarray}
where all the partial derivatives are numerically calculated from finite differences.

Now let us turn to the definition of the grandcanonical ensemble.
The standard  statistical definition of the grandcanonical partition sum, which in principle should be valid in any model, reads (we note the fugacity $\alpha=\beta\mu$):
\begin{equation}
Z_{\alpha}=\sum_{N=0}^{N_{max}(V)} Q_N \exp \alpha N = Q_0 + \sum_{N=1}^{N_{max}(V)} Q_N \exp \alpha N  \label{stat}
\end{equation}
Where $Q_N$ is the canonical partition sum corresponding to the same volume, the same temperature, and $N$ particles. In our model, we can access all $Q_N$ via the recursion relation, provided
$1\le N \le N_{max}(V)$.
Going towards the thermodynamic limit implies $V\to\infty$, and also $ N_{max}(V)\to \infty$. This means that in principle we can calculate via recursion all $Q_N$ whatever is $N$, even if of course increasing $N$ will become numerically more and more expensive.

The only unknown is then the vacuum partition sum $Q_0$. As we have discussed in ref.\cite{our_plb}, if we have an alternative expression for
$Z_\alpha$ we can deduce this unknown from the condition of normalization of probabilities:
\begin{equation}
1=\sum_{N=0}^\infty P_\alpha(N)=\frac{1}{Z_\alpha} \sum_{N=0}^\infty Q_N \exp \alpha N = \frac{1}{Z_\alpha}Q_0 + \frac{1}{Z_\alpha}\sum_{N=1}^\infty Q_N \exp \alpha N  \label{norm}
\end{equation}

Here we want to work in a finite system and not at the thermodynamical limit. Then $V$ is a fixed finite number, and
if $N>N_{max}(V)$, $Q_N=0$ because we cannot fit the particles in the finite volume. Again,  if we have an alternative expression for
$Z_\alpha$, we can deduce $Q_0$ from the condition of normalization of probabilities (\ref{norm}).

It is customary \cite{subal,our_plb} to utilize this alternative expression for $Z_\alpha$ in  the (Grand Canonical) Canonical Thermodynamical Model [(G)CTM] using cluster multiplicities. Indeed if we replace in eq.(\ref{stat}) the CTM expression for the canonical partition sum we get
\begin{equation}
 Z_{\alpha}=\sum_{N=0}^\infty \sum_{\vec{n}:N} \prod_{s=1}^{s_{max}} \frac{\omega_s^{n_s}}{n_s!}\exp \alpha N
\end{equation}
which can be rewritten as
 \begin{eqnarray}
Z_{\alpha}&=&\sum_{n_1=0}^\infty \frac{\omega_1^{n_1}}{n_1!}  \dots \sum_{n_ {s_{max}}=0}^\infty \frac{\omega_{s_{max}}^{n_{s_{max}}}}{n_{s_{max}}!} \exp \alpha N  \label{zgc1}\\
&=&\prod_{s=1}^{s_{max}}\sum_{n_s=0}^\infty  \frac{\omega_s^{n_s}}{n_s!}\exp \left ( \alpha \sum_{s=1}^{s_{max}} n_s s \right ) \label{zgc2} \nonumber \\
&=& \prod_{s=1}^{s_{max}}\sum_{n_s=0}^\infty  \frac{ \left( \omega_s
\exp\beta\mu_s \right )^{n_s}}{n_s!} \nonumber\\
&=&  \prod_{s=1}^{s_{max}} \exp \left ( \omega_s
\exp\beta\mu_s \right ) \label{zgc}
\end{eqnarray}
which is the standard expression of the grandcanonical partition sum  $Z_{GC}$ (where $\mu_s=\mu\cdot s$) as given in the literature\cite{subal}.

In this equation, $s_{max}=N_{max}(V)$. If we consider the thermodynamic limit, everything is coherent and correct.  The canonical and grandcanonical partition sums  satisfy the general relation eq.(\ref{stat}),  the quantity $P_\alpha (N)$ defined by equation (\ref{norm}) can indeed be interpreted as a probability, and we can plug eq.(\ref{zgc}) into eq.(\ref{norm})
in order to get $Q_0$.

Expression (\ref{zgc}) has been very often used in the literature.
Its great advantage from the computation viewpoint is that
the expression of cluster multiplicities is extremely simple:
\begin{equation}
\langle n_s\rangle_{\beta\mu}=\omega_s \exp\beta\mu_s
\end{equation}
and all thermodynamic quantities are defined in terms of cluster multiplicities:
\begin{eqnarray}
\beta p_{\beta\mu} &=& \frac{\partial \ln Z_{\alpha}}{ \partial V} |_{\beta\mu}=
 T \frac{ \sum_s \langle n_s\rangle_{\beta\mu}}{V}  \label{pgc1} \\
E_{\beta\mu}&=&-\frac{\partial \ln Z_{\alpha}}{\partial \beta } |_{V\mu} =
 \sum_s \langle n_s\rangle_{\beta\mu} E(s)\\
N_{\beta\mu}&=&\frac{\partial \ln Z_{\alpha}}{\partial \beta\mu } |_{V \beta}
= \sum_s \langle n_s\rangle_{\beta\mu} s
\end{eqnarray}
meaning that the model is fully analytical and no numerical calculation is needed.

A problem however arises if we work out of the thermodynamic limit and $V$ is a finite volume
where no more than $N_{max}$ particles can be fitted.
This means that the sums over multiplicities in eqs.(\ref{zgc1}),(\ref{zgc}) should not go up to infinity, but only to a maximum finite multiplicity $n_s^{max}$. For instance the last sum in eq.(\ref{zgc1}) should have only two terms, $n_{s_{max}}=0$ and $n_{s_{max}}=1$, because at most one fragment of that size filling the whole available volume
can be found.

Eq.(\ref{zgc1}) should then be transformed to:
\begin{equation}
Z_{\alpha}=\sum_{n_1=0}^{N_{max}(V)} \frac{\omega_1^{n_1}}{n_1!}  \dots \sum_{n_ {s_{max}}=0}^1 \frac{\omega_{s_{max}}^{n_{s_{max}}}}{n_{s_{max}}!} \exp \alpha N  \label{truncated}
\end{equation}

But if we introduce upper bounds on these sums 
we lose the simple expression eq.(\ref{zgc}). 

This means that in (G)CTM the grandcanonical model (\ref{zgc})  fulfils the thermodynamic relation (\ref{stat}) only in the thermodynamic limit.  When applied to a finite system, it should always be seen as an approximation of the canonical model. In particular it does not correctly describe the equilibrium of a finite system with an external particle bath,  which in principle a grandcanonical model is supposed to do.

Out of the thermodynamic limit, two situations still exist where the grandcanonical model is correct. The first situation concerns systems with Coulomb and volumes which are big enough such that the statistical weight of partitions with $N>N_{max}(V)$ is negligible. In this case we can safely make the approximation

\begin{equation}
\sum_{n_s=0}^{n_s^{max}} \frac{\omega_1^{n_s}}{n_s!}\approx \sum_{n_s=0}^\infty \frac{\omega_1^{n_s}}{n_s!}
\end{equation}

and we can recover eq.(\ref{zgc}).

 The second possibility is to consider a model without excluded volume. If we can pile up an arbitrary number of fragments in the finite volume, we will still have a maximum cluster mass $s_{max}=N_{max}(V)$ because of the finite volume, but we do not have any more a maximum multiplicity in the calculation of the grandcanonical partition sum, meaning that the upper limit in the sums of (\ref{zgc1}) is $n_s^{max}=\infty$ and (\ref{zgc}) is again correct.
In this case, if we want to calculate the grandcanonical partition sum out of eq.(\ref{stat}), we should also calculate the sum up to convergence, because without excluded volume there is no upper limit on the maximum number of particles even in a finite volume.
This is however not a satisfactory solution. Indeed
from a computation viewpoint, eq.(\ref{stat}) cannot be used because we know that the convergence is very slow\cite{fisher_noi}.
From the physical point of view, excluded volume is important and we can conclude that the grancanonical model eq.(\ref{zgc}) has to be seen only as an approximation to the grandcanonical thermodynamics, allowing infinite multiplicity of any cluster size even in a finite volume. In particular, the grandcanonical  particle number distribution
\begin{equation}
P_\alpha(N)=  \frac{1}{Z_{GC}} Q_N \exp \alpha N  \label{palpha}
\end{equation}
with $Z_{GC}$ given by eq.(\ref{zgc}), is not normalized. Still we will employ in the following
eq.(\ref{zgc}) to get an approximate estimation of the grandcanonical partition sum,
and fix the relation between $\mu$ and $N$ giving the mapping between the grandcanonical and canonical ensemble in the standard\cite{subal} way, that is solving  the equation

\begin{equation}
N=\sum_{s=1}^{N_{max}(V)}s  \exp \left ( \omega_s
\exp\beta\mu_s \right )\left ( \sum_{s=1}^{N_{max}(V)} \exp \left ( \omega_s
\exp\beta\mu_s \right ) \right )^{-1} \label{mapping}
\end{equation}

\section{Phase diagram of the neutral system}

The neutral system is a simple fluid which should exhibit a liquid-gas phase transition.
The order parameter of this transition is the particle density.  At the thermodynamic limit, a variation of the particle density is obtained by varying both the volume and the particle number
in such a way that their ratio stays finite even when $N$ and $V$ diverge.
In a finite system, a variation of the particle density can be obtained in two different ways, namely changing the number of particles in a given finite volume, or changing the volume in which a given finite number of particles is considered. Because of the different finite size effects, these different procedures are not expected to be necessarily equivalent. If we consider the dense phase where a single cluster occupies the whole volume, to have at constant volume the same energetic properties obtained at constant particle number for a system of size $N_0$, we should consider a volume $V_0$ such that $N_{max}(V_0)=N_0$.
In this volume, a unique partition is possible for any density below $1/V_0$, while a huge number of different partitions including bound clusters can be considered for the system of size $N_0$ at the same density  obtained with a volume $V=N_0V_0$.

In the following we therefore separately consider the phase diagram of a system composed of a finite number of particles, and of a system of particles contained in a finite box.

\subsection{Finite volume}

\begin{figure} [ht]
\includegraphics[width=16.0cm,keepaspectratio=true,clip]{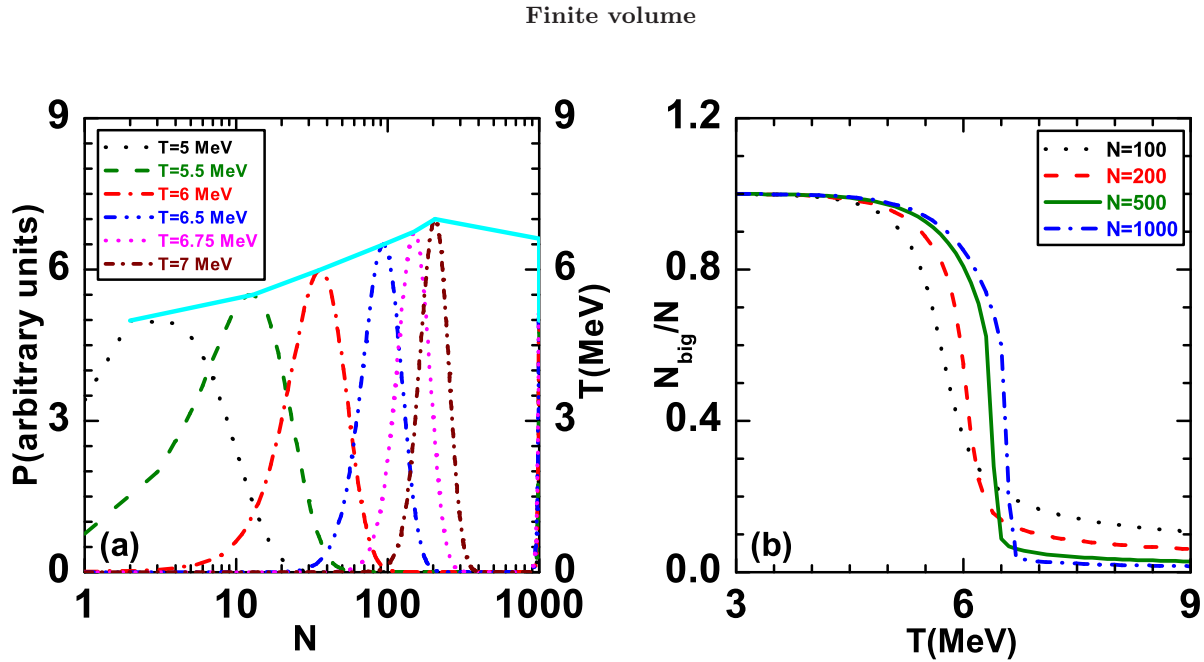}
\caption{  Left side: grandcanonical particle number distributions at different temperatures. The thick solid line gives the temperature border of the first order phase transition. Right side: average size of the largest cluster, normalized to the total particle number, as a function of the temperature in the canonical ensemble at fixed volume for different sizes of the system. Calculations are done at a volume $V=5V_0(200)$ (see text).}
\label{distri}
\end{figure}

In the last section we have shown that, for a neutral finite system, the grandcanonical model as defined by the standard expression eq.(\ref{zgc}) is not consistent. To derive the phase diagram of the model, we will therefore rely on eq.(\ref{stat}) which is always correct. We use the same technique as in ref.\cite{lehaut}. In the absence of Coulomb, the transition belongs to the liquid-gas universality class with particle number as an order parameter. In this situation, according to the general definitions of phase transitions in finite systems\cite{binder,gross}, at each temperature the transition chemical potential is determined by the condition that the grandcanonical particle number probability presents two peaks of equal height. In the cluster model, due to the hypothesis of incompressible clusters, the high density peak is always located at $N=N_{max}(V)$ \cite{fisher_noi,our_plb}. Noting $N_1(T,V)$ the particle number corresponding to the low density peak, the transition chemical potential is then given by the analytical relation $\beta\mu_t (V)=(\ln Q_N(N=N_{max}(V))-\ln Q_N(N_1))/(N_{max}(V)-N_1)$. The corresponding grandcanonical particle number distributions are shown for different temperatures and a volume $V=5V_0(200)$ (where $V_0(200)$ is the volume of a nucleus containing $200$ particles at the normal nuclear density) arbitrarily chosen such that $N_{max}(V)=1000$ in the left part of Fig.\ref{distri}.
We can see that for all the considered temperatures the low density phase is peaked at a particle number $N_1>0$. The problem, discussed in the last section, concerning the evaluation of the vacuum partition sum $Q_0$ thus does not deform the shape of the distribution and the extraction of $\mu_t$, and only  affects the normalization. In Fig.\ref{distri} we have arbitrarily fixed the undetermined normalization factor by requiring  the peaks height be numerically equal to the temperature. As a consequence, the thick line joining the probability maxima in Fig.\ref{distri} represents the phase diagram of the (G)CTM model in the temperature-particle number plane for our choice of volume. We can see that a large portion of the phase diagram corresponds to the coexistence region
of the liquid-gas phase transition. From the phenomenological point of view, this region is defined by characteristic U-shaped cluster distributions where a huge bound cluster dominates the global energetics, while the supercritical region above the thick line corresponds to exponential distributions dominated by monomers. It is however important to stress that the U-shaped mass distribution is in no way a proof of coexistence. U-shaped distributions are systematically found in the inner crust of proto-neutron stars\cite{noi,hempel}, and they do not correspond to a first order phase transition, but to a macroscopic phase with local inhomogeneities due to the phenomenon of Coulomb frustration. Other counter-examples can be found in \cite{bugaev_bimo}. The indication of the phase transition comes from the bimodality of the order parameter,  as shown in the left part of Fig.\ref{distri}.

An alternative order parameter is the size of the largest cluster. Its behavior with temperature
normalized to the total particle number is shown in  the right side of Fig.\ref{distri} for the canonical ensemble.  We can see that the largest cluster behaves as a percolating cluster exhausting most of the available mass inside the coexistence region. This is another typical signal associated to the liquid-gas phase transition. In particular, the ending point of the coexistence zone is precisely evidenced by the sudden drop of $N_{big}/N$, but this signal can only be exploited for relatively large particle numbers or the order of $N=1000$.
For smaller systems finite size effects lead to the well-known rounding of the phase transition.
It is however important to stress that this latter  can still be spotted through the bimodality signal.

\begin{figure} [ht]
\includegraphics[width=16.0cm,keepaspectratio=true,clip]{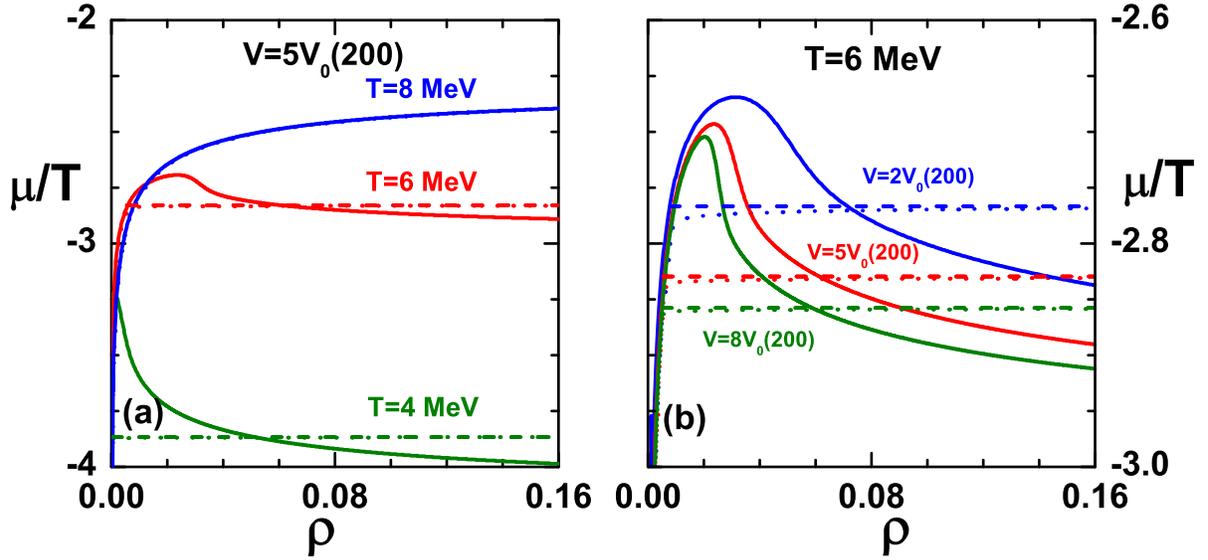}
\caption{    Left part: canonical chemical potential divided by the temperature as a function of particle density for the same temperature and volume conditions as in Fig.\ref{distri}. Solid lines: canonical results. Dotted  lines: grandcanonical results. Dashed line: coexistence plateau from Fig.\ref{distri} above.
Right part: same as the left part, but for a fixed temperature and different values for the volume.
}
\label{mu_N}
\end{figure}

 The phase diagram shown in the left part of Fig.\ref{distri} was obtained using the grandcanonical information of the particle number distribution. However this information is consistent with the finding
in the right part of the same figure, which was computed in the canonical ensemble. Indeed the
phase diagram is representative of the model and not of the specific ensemble.
A proof of this statement comes from the fact that the same phase diagram information can be extracted from the study of the canonical chemical potential\cite{lattice_prl}.
Indeed using the definition of the canonical chemical potential eq.(\ref{mucan}) and eq.(\ref{palpha}) we get
\begin{equation}
\mu_\beta= - T \frac{\partial \ln P_{\alpha}}{\partial N} + \mu
\end{equation}
Integrating over N between two arbitrary points $N_1,N_2$ gives
\begin{equation}
\int_{N_1}^{N_2}\mu_\beta(N) = - T ( \ln P_{\alpha}(N_2)- \ln P_{\alpha}(N_1)) + \mu  (N_2-N_1) \label{mucgc}
\end{equation}
Now let us choose, for the given $T, V$, the grandcanonical chemical potential $\mu=\mu_t$ that corresponds to a $p_{\beta\mu}(N)$ with two peaks of equal height. We call the particle numbers associated to the two peaks
$N_1,N_2$ and apply eq.  (\ref{mucgc}) to get
\begin{equation}
\int_{N_1}^{N_2}\mu_\beta(N) =   \mu_t  (N_2-N_1) \label{maxwell}
\end{equation}
We can identify this situation as the phase transition. Between the beginning $N_1$ and the end $N_2$ of the coexistence region, the canonical backbending chemical potential fulfills an equal area Maxwell construction (\ref{maxwell}).
The points $N_1$ and $N_2$ represent the particle numbers associated to the dilute and dense phase respectively, which are equally probable at the transition grandcanonical chemical potential $\mu_t$.\\
The pertinent equation of state at fixed volume is the functional relation between particle number and chemical potential. This equation of state is represented for the two ensembles in Fig.\ref{mu_N}, in the same thermodynamic conditions as before. As it is well known, while the grancanonical equation of state is monotonous, the canonical one shows an inversed (decreasing) behavior which is the typical signal of the phase transition. The verification  that eq. (\ref{maxwell}) holds in the numerical calculations of the (G)CTM is also given in the same figure. The canonical chemical potential from eq.(\ref{mucan}) is represented as a function of the particle number at a given volume and for different temperatures, together with the value corresponding to the equal area construction eq.(\ref{maxwell}). We can see that this phase transition definition leads to the same phase diagram as in Fig.\ref{distri}. The consistency of the phase diagram of the cluster model calculated from different statistical ensembles was not verified before to our knowledge. The effect of the volume on the phase diagram is explored in the right part of Fig.\ref{mu_N}. We can see that the qualitative behavior of the phase diagram is independent of the volume. At a given temperature, decreasing the volume has only a very small effect on the density width of the coexistence zone $(N_2-N_1)/N_2$, but leads to an increase of the transition chemical potential. This result is consistent with the liquid-gas transition phenomenology of simple fluids, while the small sensitivity of the density transition point with the volume might be  attributed to our simplifying hypothesis of incompressible clusters.\\
\begin{figure} [ht]
\includegraphics[width=16.0cm,keepaspectratio=true,clip]{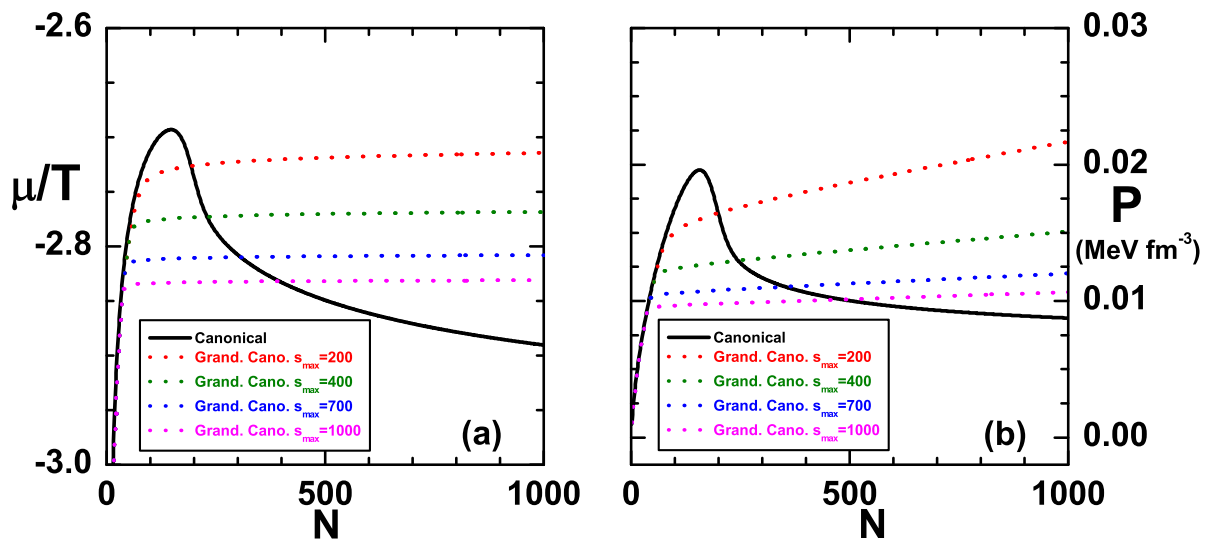}
\caption{  Left part: canonical (black line) and grandcanonical (colors online) chemical potential at $T=6$ MeV for the same volume as in Fig.\ref{distri}. The different grandcanonical curves correspond to different maximum cluster sizes allowed in the volume $V=5V_0$, with $N_{max}(V)=1000$. Right part: same as for the left part, but for the pressure.}
\label{mu_kmax}
\end{figure}
The dashed lines in Fig.\ref{mu_N} give the grandcanonical results, where the correspondence $\langle N \rangle_{\beta\mu}(\mu)$ is given by eq.(\ref{mapping}). Coherently with the general theory of phase transitions in finite systems\cite{gross,lattice_prl}, the grandcanonical chemical potential is monotonous and very close to a plateau in the density region corresponding to the backbending. Previous studies have shown\cite{subal,gargi07} that a plateau is exactly recovered in the thermodynamic limit, as expected for the liquid-gas phase transition of a simple fluid. It is interesting to remark that the chemical potential value of the plateau is very close to the grandcanonical transition chemical potential defined by eq.(\ref{maxwell}), in spite of the fact that eq.(\ref{mapping}) uses the approximate expression $Z_{GC}$ from eq.(\ref{zgc}) to  evaluate the grandcanonical partition sum. This   result implies that the incorrect treatment of excluded volume does not affect the phase properties of the system in an important way.
\\
In previous works\cite{subal,gargi07}, the possibility of introducing a maximum cluster size $s_{max}$ in the grandcanonical ensemble was discussed. In these works, it was already shown that the thermodynamic properties of the model strongly depend on the choice of this parameter. The effect on the chemical potential of modifying the maximum cluster size in the grandcanonical  ensemble is shown in the left part of  Fig.\ref{mu_kmax} for a representative value of volume and temperature. We can see that indeed the transition chemical potential wildly changes by modifying $s_{max}$. This is easy to understand, because a partition containing a single cluster of size $s_{max}$ is the best finite system approximation of the dense liquid phase. Changing $s_{max}$ thus amounts to artificially change, for a fixed finite size given by the finite volume, the finite size effects on the phase transition. Fig.\ref{mu_kmax} shows that $s_{max}$ is not a free parameter, and it has to be fixed as $s_{max}=N_{max}(V)$ in order to have thermodynamically consistent results in the (G)CTM model.\\
\begin{figure} [ht]
\includegraphics[width=16.0cm,keepaspectratio=true,clip]{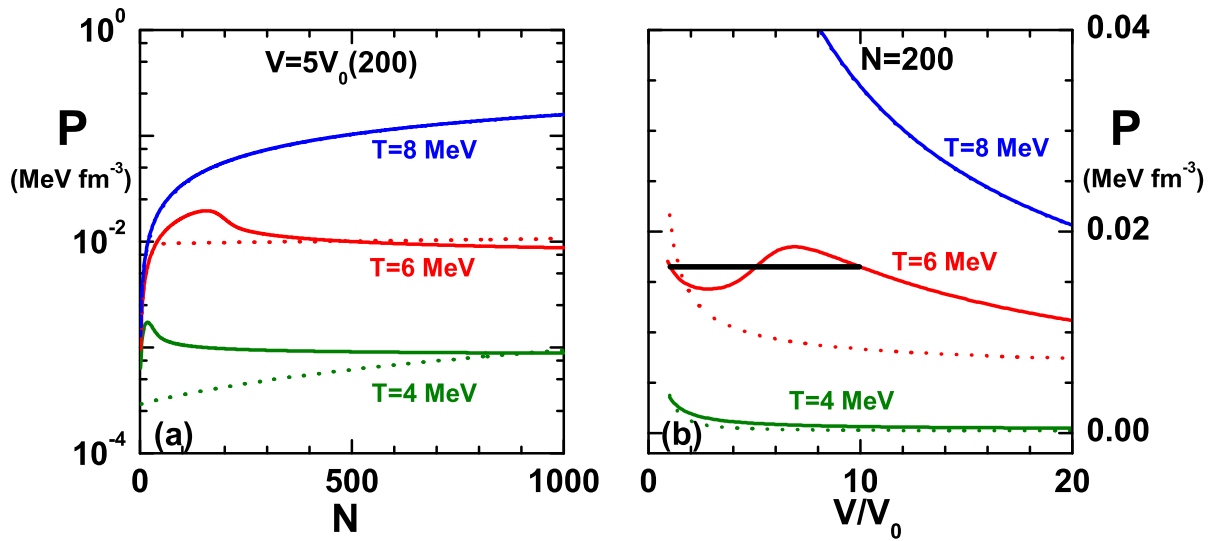}
\caption{ Canonical (full lines) and grandcanonical (dashed lines) pressure as a function of (i) the number of particle for a fixed volume $V=5V_0(200)$ (left column) and (ii) the volume for a system of $N=200$ uncharged particles (right column). In the right column, for $T=6$ MeV, the thick line represents the Maxwell equal area construction.
}
\label{p_V}
\end{figure}
The behavior of the canonical and grandcanonical pressure from eqs.(\ref{pc1}),(\ref{pgc1}) is shown in the right part of Fig.\ref{mu_kmax}, in the same thermodynamic conditions as for the left part of the figure.  Again, the choice of the maximum cluster size has a deep influence on the grandcanonical equation of state and only the result obtained fixing $s_{max}=N_{max}(V)$ should be considered as physically meaningful and thermodynamically consistent.
We can see that, similar to the behavior of the chemical potential, the two ensembles are equivalent for very low particle numbers, where according to the phase diagram Fig.\ref{distri} the system is in a pure gas phase. If $s_{max}=N_{max}(V)$, the point where ensemble inequivalence starts to appear perfectly coincides with the border of the coexistence line, showing again  that the approximate expression $Z_{GC}$ from eq.(\ref{zgc}) is good enough to determine the phase properties of the system. Also, the spinodal region can be equivalently defined from the presence of negative susceptibility ($\partial \mu_\beta/\partial N<0$) or negative compressibility ($\partial P_\beta/\partial N<0$)
\subsection{Finite particle number}
In the previous figures we have analyzed the phase transition by considering a fixed finite box which could be filled with a different number of particles. This situation can be physically relevant for metallic nanoparticles in rigid structures\cite{fabienne}. In the case of fragmentation of atomic nuclei or atomic clusters, the finiteness of the system is rather manifested by the fact that a finite and fixed number of particles can occupy different volumes (for instance because it evolves in the vacuum following an excitation process). In this case the order parameter is volume and the equation of state to be analyzed is $P(V)$. This situation  has been already studied many times in the past\cite{subal,gargi_fra,gargi07}. A representative calculation is presented in Fig.\ref{p_V} for a system of $N=200$ neutral particles. The corresponding $P(N)$ diagram for fixed volume $V=5V_0$ is plotted in the left panel for the same value of temperatures for both the ensembles.\\
Though still very much evocative of the liquid-gas phase transition of simple fluids at the thermodynamical limit, the behavior of the isotherm is deeply different from the observations at fixed volume. The canonical and grandcanonical calculations only converge at high temperature, in the supercritical region. The grandcanonical pressure presents no plateau and is always monotonically decreasing as expected in an ideal gas. This is expected. The order parameter being the volume, a plateau can only be found in the canonical isobar ensemble, that is by fixing the particle number and letting the volume fluctuate with the introduction of a conjugated pressure field.
This has been done in the Monte-Carlo microcanonical version of the cluster model in ref.\cite{raduta}.
Concerning the canonical calculations,
the low temperature isotherms appear monotonously decreasing, indicating a pure liquid-like phase independent of the volume. At higher temperature a negative compressibility appears, signalling the finite system counterpart of the phase coexistence. The phase diagram can be obtained with the same procedure that we have followed for the constant volume case. The canonical isobar ensemble is defined by the partition sum
\begin{equation}
Z_{\beta p}=\sum_{V=V_0(N)}^{\infty} Q_N(V) \exp \beta p V  \label{isobar}
\end{equation}
and the volume probability distribution is given by:
\begin{equation}
P_{\beta p}=Z_{\beta p}^{-1} Q_N(V) \exp \beta p V
\end{equation}
If at a given temperature the canonical pressure backbends, this means that in the isobar canonical ensemble the volume probability distribution must present two peaks, located at two values $V_1(T)$, $V_2(T)$. These peaks are equally probable at a given pressure $p_t(T)$. With the same reasoning as in the previous section, we can define the coexistence zone from an equal area construction of the canonical equation of state, according to:
\begin{equation}
\int_{V_1}^{V_2}p_\beta(N) =   p_t  (V_2-V_1) \label{maxwell2}
\end{equation}
which can be used to define $p_t$. This construction, defining the borders of the coexistence zone, is shown in the right column of Fig.\ref{p_V} by the thick line. We can see that at moderate temperatures the coexistence zone covers a large part of the phase diagram, as we have seen in the fixed volume case. Still the smallest volumes (see for instance the isotherm at $T=5.5$ MeV in the left panel of Fig. 1 ) correspond to a pure liquid phase. If we consider for instance the point $T=4, V=5V_0, N=200$, such a thermodynamic situation, which is characterized by a U-shaped cluster size distribution, is interpreted as a coexistence in the phase diagram of $V=5V_0$, and as a pure liquid phase in the phase diagram of $N=200$.\\
This example is a nice illustration of the importance of working in the physically meaningful statistical ensemble when studying the statistical mechanics of finite systems.\\
\section{Phase diagram of the charged system}
We now turn to examine the complete model where the long range Coulomb interactions are approximately accounted in the Wigner Seitz approximation.\\
\begin{figure} [ht]
\includegraphics[width=16.0cm,keepaspectratio=true,clip]{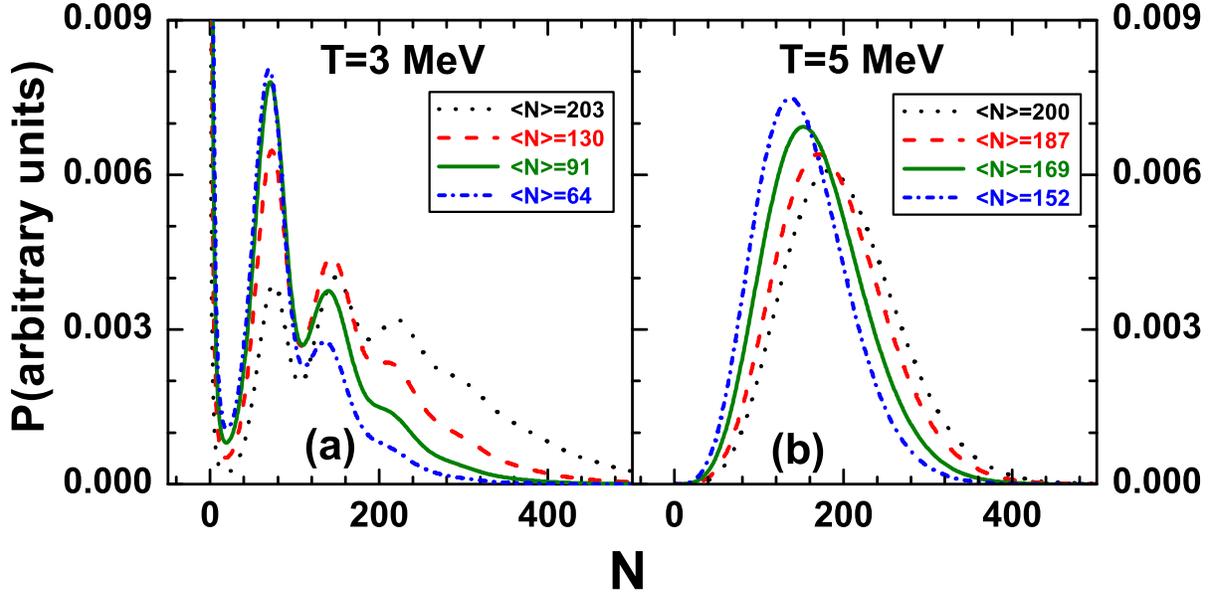}
\caption{   Particle number distribution in the grand-canonical ensemble for T=3 MeV (left) and
T=5 MeV (right), with the same volume as in Fig.\ref{distri} above and different average particle numbers, given in the figure.}
\label{distri_coul}
\end{figure}
The particle number distributions eq.(\ref{palpha}) at different temperatures and chemical potentials are displayed in Fig.\ref{distri_coul}, to be compared to the similar analysis for the uncharged system in the left panel of Fig.\ref{distri}.\\
The first, relatively trivial effect of the inclusion of the Coulomb interaction is that normal distributions, typical of a pure phase, are obtained at temperatures as low as 5 MeV, a temperature which still corresponds to the coexistence region in the uncharged system (see Figure \ref{distri}).
These distributions are not pure gaussians as it is typically expected from finite size effects because of the border effect given by the fact that a particle number cannot extend below zero. Still, they show an approximate gaussian behavior as it is seen in the uncharged system above the thermodynamical critical point, with a width increasing with the temperature following the thermodynamical relation
which relates the susceptibility $\chi$ to the particle number fluctuation $\sigma^2$: $\chi=\partial N_{\beta\mu}  /\partial \mu = \beta \sigma_N^2$.
This quenching of the phase transition, with a reduction of the critical temperature, is expected due to the repulsive character of the Coulomb interaction.\\
However the effect of  Coulomb is much more drastic than a simple shift of the phase diagram, as shown in the left part of Fig.\ref{distri_coul}. The purely bimodal behavior of the distribution is not recovered at any temperature, and is replaced by a complex structure showing different peaks with approximately equal spacing. As it was already observed in ref.\cite{our_plb}, these peaks indicate preferential fragmentation structures corresponding to higher multiplicity of clusters of similar size.\\
Still, up to around $T\approx 4 MeV$, a transition chemical potential $\mu_t(T)$ can be defined such that the probability associated to the vapor ($N=1$)
is equal to the probability associated to the droplet ($N\approx 200$).
This can be easily understood from the fact that, due to Coulomb, the liquid phase apparent in Fig.\ref{distri} is replaced by a phase constituted of clusters.\\
\begin{figure} [ht]
\includegraphics[width=16.0cm,keepaspectratio=true,clip]{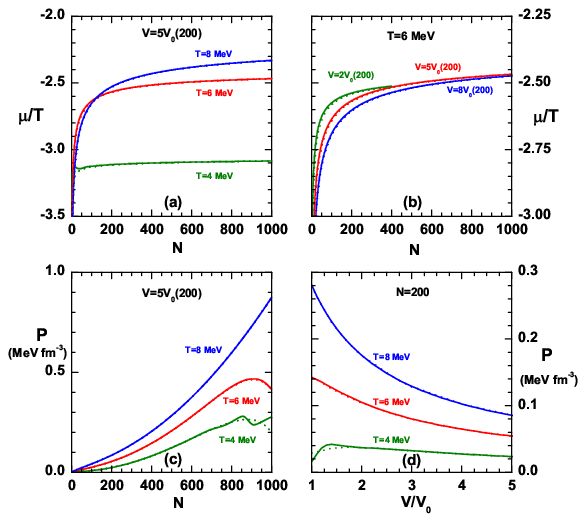}
\caption{Thermodynamics of a charged system. Canonical and grandcanonical chemical potential (upper part) and pressure (lower part) as a function of the particle number at fixed value, and as a function of the volume for a fixed particle number (lower right).  Different temperatures are considered.}
\label{p_mu_coul}
\end{figure}
The model with Coulomb does not admit a thermodynamical limit, therefore it is not possible to study the behavior for $N$ or $V\to\infty$. However it is easy to verify  that an increase of the  total volume does not change the structure of the two phases, but simply leads to a displacement of the second peak towards higher particle numbers, corresponding to a higher number of clusters in the dense phase.\\
Because of this persistence of phase transition signals in the presence of Coulomb, we can expect inequivalence signals to arise in the different equations of state, similar to the neutral case studied in the previous chapter. This is confirmed by Fig.\ref{p_mu_coul}, which displays the behavior of chemical potential and pressure in the same thermodynamic conditions explored for the neutral system. We can appreciate the considerable quenching of the phase transition at high temperature. Still, a region of ensemble inequivalence and inversed slope for the canonical equations of state persists at low temperature and high density, reminiscent of the liquid-gas phenomenology enlighten above.\\
In order to show that we are still facing a coexistence phenomenon between a vapor of monomers and (charged) droplets,
we show in Fig.\ref{big_coul} the distributions of the heaviest cluster\cite{subal} associated to each fragmented configuration in the ensemble inequivalence region. We can see that the grandcanonical distribution, even when very high particle numbers are implied, is always characterized by two peaks, corresponding to the monomer solution and the droplet solution. The mass of the largest cluster in the denser phase obviously depends on the available volume, but is always upper limited because of the repulsive Coulomb force.\\

\begin{figure} [ht]
\includegraphics[width=16.0cm,keepaspectratio=true,clip]{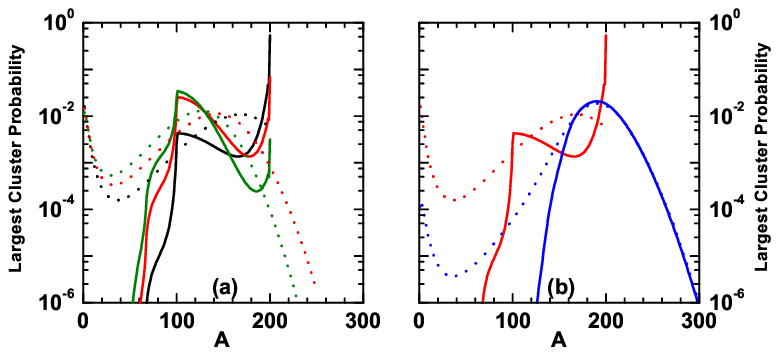}
\caption{Distribution of the largest fragment associated to each fragmented configuration in the ensemble inequivalence region in the grandcanonical (dotted lines) and canonical (full lines) ensemble. Left side: the total number of particles (average number for the GC calculation) is fixed to $N=200$ and three different volumes $2V(200)$ (black), $2.25V(200)$ (red) and $2.5V(200)$ (green) are considered. Right part: two different (average in the GC case) number of particles $N=200$ (red) and $N=1000$ (blue) are considered and the volume is fixed such that $V=2.25 V_0(N)$. Calculations are done at a fixed temperature $T=4$ MeV.}
\label{big_coul}
\end{figure}

The interesting point of this figure is that the low $A_{max}$ peak is always located at $A_{max}=N=1$, that is it corresponds to much lower density
than the high $A_{max}$ peak. Density can therefore be viewed as the order parameter of the transition even in the presence of Coulomb.
This original result means that, in the framework of this model, fragmentation can still be viewed as a liquid-gas type phase transition even in the presence of Coulomb.
When the mass conservation constraint is implemented at the same time as the volume conservation constraint (canonical ensemble) the vapor solution becomes inaccessible because a much higher volume would be needed in order to reach its low density. As a consequence, a complex distribution
with two peaks corresponding to different heaviest cluster sizes are observed, as we have already pointed out in our previous work\cite{gargi_fra}.
We can however observe that, increasing the particle number such that the conservation constraint becomes less strict (right part of Fig.\ref{big_coul}),
the distribution becomes single-peaked again, corresponding to fragmentation configurations where multiple droplets of the biggest size permitted by Coulomb are formed.\\
We can therefore conclude that the presence of a peak at $A_{max}\approx N/2$ in the canonical fragmentation of the small system (left part of Fig.\ref{big_coul}) is, at least  in this model and in the framework of the Wigner-Seitz approximation, an effect of particle number conservation, and not the manifestation of a new fission-like phase transition. From the nuclear physics point of view, these findings imply that the bimodality signal appears as a robust signal of the phase transition in nuclear fragmentation.
\section{conclusions}
In this paper we have studied the phase diagram of the  thermodynamical cluster model in the presence of finite size and Coulomb effects. We have shown that, even for arbitrarily small particle numbers, a phase coexistence region can be clearly identified by the behavior of the grandcanonical particle number distribution. The different phase transition signals proposed in the literature (backbendings and bimodality of the heaviest cluster distribution) are consistently found in this coexistence region. Depending on the statistical ensemble, the behavior of the equations of state and the definition itself of the different phases can be very different. To give a single example, no back-bending is observed in the $\mu(N)$ relation at low temperature if $V$ is fixed, while a clear back-bending is visible in $p(V)$ for fixed $N$. In both cases however, a transition chemical potential and pressure can be clearly be defined by the associated grandcanonical distributions. Because of this complexity, the clearest signals of phase transition are given by the simple fact that qualitative differences in the observables are obtained using different statistical ensembles. This ensemble inequivalence as a signature of phase transition persists even in the presence of the Coulomb interaction. In particular, an equal probability is observed in the grandcanonical ensemble between a vapor and a condensed solution. This condensed solution is given by a phase characterized by a number of the heaviest clusters that Coulomb can sustain, the number of clusters being essentially determined by the available total particle number and/or volume.\\
This discussion on the implications of the phase transition in the different ensembles is a somewhat academic discussion for the physical situations which can be experimentally accessible. Some examples exist in low-dimensional condensed matter physics\cite{fabienne} where it is possible to produce finite systems in different statistical ensembles, and it would be very interesting to apply these consideration to the physics of nanowires. However, concerning nuclear physics for which the cluster model was originally developed, the statistical ensemble can hardly be varied.
Specifically, in the standard canonical ensemble which is believed to be the most appropriate to describe the experimentally accessible situation of nuclear fragmentation, the conservation law on both the total particle number and total volume prevents from observing this bimodal behavior, but a (strongly deformed) bimodality signal persists in the distribution of the heaviest cluster, as it was reported previously\cite{gargi_fra,our_plb} Even if baryon number is exactly conserved, different traditional models of nuclear physics violate this conservation law. This is especially true in the case of pairing, where particle number conservation, violated by the BCS theory, is typically accounted for by (approximate) projection methods. In view of our results, it would be very interesting to study the finite size effects with the methods introduced in this paper, in the superfluid-normal fluid phase transition.\\
An other interesting prolongation of the present work concerns the case of stellar matter, which constitutes an intermediate case between the charged and uncharged case. Finite temperature matter formed in core-collapse supernova and in the crust of proto-neutron stars is in equilibrium with respect
to strong interactions. Its density is not fixed by a conservation law but imposed by the external gravitational pressure.
This means that, at variance with the laboratory situation, the grandcanonical ensemble is the appropriate ensemble for a statistical description of dense matter. However, most theoretical modelizations are performed in the canonical ensemble constituted by a finite Wigner-Seitz cell.
If the ensemble inequivalence we have observed is kept in the case of stellar matter, this could have interesting consequences on the composition of
the crust of neutron stars at finite temperature. This perspective is left for future work.

\end{document}